\documentclass[a4paper,oneside,11pt,reqno]{amsart}
\usepackage{amssymb,bbm}
\usepackage{amsthm}
\usepackage{amsmath}
\usepackage{cite}
\usepackage{graphicx}
\usepackage[margin=2.9cm]{geometry} 
\usepackage[foot]{amsaddr}

\usepackage{hyperref}
\hypersetup{
	pdftitle = {Dimension towers of SICs I: Some constructions}, 
	pdfauthor = {Ingemar Bengtsson, Basudha Srivastava}
}

\expandafter\def\expandafter\normalsize\expandafter{%
    \normalsize
    \setlength\abovedisplayskip{12pt}
    \setlength\belowdisplayskip{12pt}
    \setlength\abovedisplayshortskip{10pt}
    \setlength\belowdisplayshortskip{10pt}
}

\makeatletter
\renewcommand{\@secnumfont}{\bfseries}
\renewcommand\section{\@startsection{section}{1}%
 \z@{.7\linespacing\@plus\linespacing}{.5\linespacing}%
  {\indent\normalfont\bfseries}}
\makeatother

\newtheorem{theorem}{Theorem}

\begin{document}

\vspace{6cm}

\begin{center}

\

\

{\Large DIMENSION TOWERS OF SICS. II.}

\

\

{\Large SOME CONSTRUCTIONS}

\vspace{14mm}

{\large Ingemar Bengtsson}$^{*}$

\vspace{7mm}

{\large Basudha Srivastava}$^{\dagger}$

\vspace{14mm}

$^{*}${\small {\sl Stockholms Universitet, AlbaNova, Fysikum, \\
SE-10691 Stockholm, Sverige}}

\

${}^{\dagger}${\sl {\sl G\"oteborgs Universitet, Institutionen f\"or fysik, \\
SE-41296 G\"oteborg, Sverige}}

\vspace{14mm}

{\bf Abstract:}

\vspace{7mm} 

\parbox{112mm}{
{\small \noindent A SIC is a maximal equiangular tight frame in a finite dimensional 
Hilbert space. Given a SIC in dimension $d$, there is good evidence that there always 
exists an aligned SIC in dimension $d(d-2)$, having predictable symmetries and smaller 
equiangular tight frames embedded in them. We provide a recipe for how to calculate sets 
of vectors in dimension $d(d-2)$ that share these properties. They consist of maximally 
entangled vectors in certain subspaces defined by the numbers entering the $d$ dimensional 
SIC. However, the construction contains free parameters and we have not proven that 
they can always be chosen so that one of these sets of vectors is a SIC. We give 
some worked examples that, we hope, may suggest to the reader how our 
construction can be improved. For simplicity we restrict ourselves to the case 
of odd dimensions.}} 

\end{center}

\newpage

\section{Introduction}\label{sec:intro}

\

\noindent A tight frame in a complex $d$ dimensional Hilbert space is a collection of 
$N$ unit vectors that provide a resolution of the identity. In quantum information 
theory tight frames are known as POVMs. If in addition we require the absolute 
values of all the mutual overlaps between the vectors to be the same, then the tight frame 
is said to be equiangular (and the POVM is said to be symmetric). Translating this 
into equations one easily finds that the vectors $|\psi_I \rangle$ in an equiangular 
tight frame (abbreviated ETF) must obey 

\begin{equation} \sum_{I=1}^N|\psi_I\rangle \langle \psi_I| = \frac{N}{d}{\bf 1} \ , 
\hspace{8mm} |\langle \psi_I|\psi_J\rangle|^2 = \left\{ \begin{array}{ccl} 1 & \mbox{if} 
& I = J \\ \frac{N-d}{d(N-1)} & \mbox{if} & I \neq J \ . \end{array} \right. 
\label{etfs} \end{equation}

\noindent It is also easy to prove that $d\leq N \leq d^2$, but it is not at all easy 
to see for what numbers $N$ of vectors an  ETF exists \cite{Mixon}. When written as 
multivariate polynomial equations for the components of the vectors the defining 
equations have a distinctly intimidating look. Anyway the minimal case $N = d$ is that 
of an orthonormal basis, which is a very important ETF that exists for any $d$. It 
takes only a small amount of curiousity to ask whether the maximal case $N = d^2$ is 
important too. Maximal ETFs are known as SICs---which we treat as a name, 
not as an abbreviation of anything. 

It was conjectured, by the authors who first brought them to the attention of the 
world, that SICs do exist in all dimensions. Moreover they were conjectured to exist 
as orbits under the Weyl--Heisenberg group \cite{Zauner, Renes}. Since this is a 
group with a very large number of applications in quantum theory and in classical 
signal processing, the conjecture does strengthen the belief that SICs will 
prove important at least in these two areas. Another early conjecture was that 
when formed in this way SICs are always left invariant by a unitary symmetry of 
order three \cite{Zauner, Marcus}. This became known as the Zauner symmetry, and 
was an early indication that something interesting is going on in the depths below. 

For the purposes of this paper it is convenient to include the requirement that 
a SIC be an orbit of the Weyl--Heisenberg group in the definition of a SIC, and we 
will do so from now on. This has the further advantage that we can fix the standard 
clock-and-shift representation of the group \cite{Weyl}. Once this has been done we 
can meaningfully ask for the kind of numbers that are needed to form the components 
of the vectors in a SIC. In this direction, a surprising addition to the existing 
conjectures was provided a few years ago. 

Based on the solutions that had been found at the time, Appleby et al. conjectured that, 
in a standard basis provided by the Weyl--Heisenberg group, the number field generated by 
a SIC in dimension $d$ is, or includes, a ray class field 
over a real quadratic base field with conductor $d$ \cite{AFMY}. Without going into 
detailed explanations we observe that the problem of finding generators for these 
number fields was raised by Hilbert in his so far unsolved 12th problem 
\cite{Hilbert}. This suggests another angle from which SIC existence is important: 
to borrow a phrase, it ``reshapes the boundary between physics and pure mathematics'' 
\cite{John}.

Judging from numerical searches and exact solutions, all these conjectures are true 
\cite{Scott, Andrew, ACFW, FHS, GS}. And the testing has been extensive: so far 
unpublished work by Grassl includes exact solutions in more than one hundred 
different dimensions \cite{MGun}, and the connection to number theory has enabled 
us to construct SICs in several four and five digit dimensions \cite{Starkpap}. 
But it is not clear how close we are to a general existence proof for SICs---and 
still less what implications a proof would have for quantum theory, signal 
processing, number theory, or the boundary between physics and pure mathematics. 

The ray class conjecture does, however, suggest an intermediate aim because it 
implies that there are infinite sequences of dimensions in which the number 
field needed to construct a SIC in one dimension is a subfield of the number field 
needed to construct a SIC in the next dimension \cite{AFMY}. Here we will focus on 
special subsequences of dimensions $\{ d_n\}_{n=0}^{\infty}$ obtained recursively by 

\begin{equation} d_{n+1} = d_n(d_n-2) \ . \label{sekvens} \end{equation}

\noindent They are called {\it dimension ladders}. We assume that $d_0 > 3$. Our 
main conjecture is that for every SIC in dimension $d_n$ there exists an {\it aligned} SIC 
in dimension $d_{n+1}$ 
such that the number field needed to construct the latter contains the number field 
needed to construct the former. Moreover the phase factors in the overlaps of the SIC 
vector pairs (which are not stipulated by the definition of a SIC) in dimension $d_n$ 
reappear in squared form among the overlap phases for the aligned SIC in dimension 
$d_{n+1}$ \cite{Gary}. 
This gained additional interest when it was observed that (Galois conjugates of) the 
squared overlap phases play a particularly interesting number theoretical role \cite{Kopp}. 
Closer investigation gave us a much more detailed (if still largely conjectural) 
picture, and provided more than twenty explicit examples of this alignment phenomenon. 
In particular it was noticed that aligned SICs in dimensions of the
form $d(d-2)$ have to contain Weyl--Heisenberg orbits of smaller equiangular tight frames 
sitting inside them \cite{aligned, AD}. 

In this paper we will go the other way: Using only information 
provided by a SIC in dimension $d$ we will explicitly construct these orbits of 
equiangular tight frames. In fact we will find a continuous family of such orbits, 
and are then left with the problem of locating the aligned SICs within them. We do 
not solve this problem in general, but we take steps in the right direction. 

We are thus led to introduce a new conjecture, which is that it should be possible to climb 
the dimension ladders rung by rung, in the sense that it should be possible to prove the 
existence of a SIC in dimension $d(d-2)$ given a SIC in dimension $d$. This would 
prove SIC existence in many infinite sequences of dimensions. However, 
while we find the evidence for all the other conjectures we have mentioned to be 
overwhelming, we are not at all convinced of the truth of this last one. Our purpose 
here is to present a few theorems to pinpoint exactly where missing ingredients are 
encountered. 

Section \ref{sec:background} provides a sketch of the group theoretical background, with some useful 
technicalities relegated to the Appendix. At the end of Section \ref{sec:background} we also recall some 
definitions from frame theory. Section \ref{sec:aligned} introduces the ladders formed by aligned SICs, 
and gives the key properties of the latter. Sections \ref{sec:theorem1} and \ref{sec:theorem2} present the 
theorems that are our main results. Given a SIC in dimension $d$ they 
allow us to calculate Weyl--Heisenberg orbits of equiangular tight frames embedded 
in dimension $d(d-2)$, and it is shown how one can impose symmetry requirements on 
these orbits. Since the details are a little involved, Section \ref{sec:recipe} condenses Sections \ref{sec:theorem1} 
and \ref{sec:theorem2} into a few easy-to-follow steps. The following sections describe explicit examples 
in some detail. We summarize our findings in Section \ref{sec:summary}, and allow the reader to 
draw her own conclusions.

\section{Background material}\label{sec:background}

\

\noindent To prevent too much overlap with previous papers we will rely freely on 
several facts concerning the Weyl--Heisenberg group, its symplectic automorphism group, 
and the Chinese remainder theorem as applied to these groups. If the reader is unfamiliar 
with these ingredients we refer to ref. \cite{aligned} for a short introduction, to 
the references therein for long ones, and to the Appendix for a few useful formulas. 
But we do need to fix some notation. First 

\begin{equation} \omega_d = e^{\frac{2\pi i}{d}} \ , \hspace{10mm} \tau_d = 
- e^{\frac{i\pi}{d}} \ . \end{equation}

\noindent Sub- and superscripts referring to the dimension will be dropped if they 
are not needed. Second, the displacement operators $D_{\bf p}$ are defined by 

\begin{equation} D_{\bf p} = D_{i,j} = \tau^{ij}X^iZ^j \ . \end{equation} 

\noindent Here $Z$ and $X$ obey $ZX = \omega XZ$ and $X^d = Z^d = {\bf 1}$. They are 
the clock and shift operators in the standard Weyl representation, which takes the 
operator $Z$ to be diagonal \cite{Weyl}. The labelling 
`vector' ${\bf p}$ has two components $i$, $j$ that are integers counted 
modulo $d$ (in odd dimensions). The displacement operators form a unitary operator 
basis acting on ${\bf C}^d$. The SIC is 
constructed by applying all of them to a fiducial vector $|\Psi_{\bf 0}\rangle$, 
and we define the SIC overlap phases by 

\begin{equation} e^{i\theta_{\bf p}} = \left\{ \begin{array}{ccl} 1 & \mbox{if} & 
{\bf p} = 0 \\ \\ \sqrt{d+1}\langle \Psi_{\bf 0} |D_{\bf p} |\Psi_{\bf 0} \rangle & 
\mbox{if} & {\bf p} \neq {\bf 0} \ . \end{array} \right. \end{equation}

\noindent They are phase factors by the definition (\ref{etfs}) of the SIC, but they 
are not fixed by the definition. Orbits of unitarily equivalent SICs are created by 
$SL(2, {\bf Z}_d)$, the symplectic group of two-by-two matrices with entries in the 
set of integers modulo $d$. Given a matrix $F \in SL(2, {\bf Z}_d)$ its unitary 
representative $U_F$ is determined up to an overall phase factor by the representation 
of the Weyl--Heisenberg group, and its matrix elements are $d$th roots of unity again 
up to a harmless normalisation factor \cite{Marcus}. This group also gives rise to 
unitary symmetries of individual SICs, while anti-symplectic matrices give rise to 
anti-unitary symmetries (if any). The symplectic group and the Weyl--Heisenberg group 
taken together generate the Clifford group. 

When the dimension is composite with relatively prime factors, $d = n_1n_2$ 
with $(n_1,n_2) = 1$, the Weyl--Heisenberg group and its symplectic automorphism 
group splits into a direct product in a canonical way. Thus, effectively,  

\begin{equation} D^{(n_1n_2)}_{\bf p} = D^{(n_1)}_{{\bf p}^\prime} \otimes 
D^{(n_2)}_{{\bf p}^{\prime \prime}} \ , \hspace{8mm} U^{(n_1n_2)}_{F} = 
U^{(n_1)}_{F^\prime} \otimes U^{(n_2)}_{F^{\prime \prime}} \ . 
\label{Kina} \end{equation}  

\noindent We use superscript to denote the dimension on which the operators act 
only when this is necessary to avoid confusion. The labelling `vectors' and matrices 
${\bf p}^\prime$, $F^\prime$, ${\bf p}^{\prime \prime}$, $F^{\prime \prime}$ 
are computed from ${\bf p}$ and $F$ through a simple application of the 
Chinese remainder theorem \cite{aligned}. 

Symplectic unitaries worth special mention are the order 2 parity operator $U_P$, 
and the order 3 Zauner unitary $U_{\mathcal{Z}}$. The parity operator is obtained from the 
$SL(2, {\bf Z}_d)$ matrix 

\begin{equation} P = \left( \begin{array}{rr} -1 & 0 \\ 0 & -1 \end{array} \right) 
\ . \end{equation} 

\noindent Its unitary representative is a permutation matrix. This is actually true 
for every diagonal symplectic matrix \cite{Marcus}. In prime 
dimensions the Zauner unitary $U_{\mathcal{Z}}$ belongs to a unique conjugacy class of the 
group. A standard choice for the corresponding $SL(2, {\bf Z}_d)$ matrix is 

\begin{equation} \mathcal{ Z} = \left( \begin{array}{rr} 0 & -1 \\ 1 & -1 \end{array} \right) 
\ . \label{Zaunermatris} \end{equation}

\noindent A point worth noticing is that whenever $d = 3$ or $d = p = 1$ modulo 3, 
where $p$ is a prime, the conjugacy class to which the Zauner matrix $\mathcal{ Z}$ belongs 
contains a representative whose corresponding unitary is a permutation matrix \cite{Marcus}. 
This is particularly relevant here because, with the possible exceptions of $d_0$ and $d_1$, 
all the dimensions in the sequences (\ref{sekvens}) are divisible by 3 so that the 
dimensions we are factoring in, of the form $d_n-2$, are equal to 1 modulo 
3. In at least one case, the ladder starting at $d_0 = 5$, this is also true for all 
the prime factors of $d_n - 2$, for all $n > 1$ \cite{IBGM}. 

We will be interested in the eigenspaces of the unitary operators $U_P$ and 
$U_{\mathcal{Z}}$ \cite{Zauner}. Their dimensions are given in Table \ref{tab:spectra}. 
The eigenvalues are convention dependent, but the sizes of the eigenspaces are not. 

\begin{table}[t]
\caption{{\small Dimensions of the eigenspaces for the parity and Zauner unitaries 
\cite{Zauner}, with $\omega_3=e^{2\pi i/3}$ and $d$ odd. An eigenspace is in boldface 
if standard conjectures \cite{Scott, Andrew} imply that it contains a SIC vector. 
The brackets that occur in one case signify that (as far as we know \cite{Andrew}) 
this eigenspace contains a SIC vector only if $d = 8$ modulo 9. The eigenvalues 
are convention dependent.}}
\vskip 5mm
{\renewcommand{\arraystretch}{1}
\begin{tabular}{|l|c|c|c|} \hline  
$U_P$ & $d$ & $1$ & $-1$ \\
\hline 
& $2n+1$ & $n+1$ & $n$ \\
\hline 
\end{tabular} 
\hskip 4mm
\begin{tabular}{|l|c|c|c|c|} \hline  
$U_{\mathcal{Z}}$ & $d$ & $1$ & $\omega_3$ & $\omega_3^2$ \\
\hline 
& $6k+1$ & ${\bf 2k+1}$ & $2k$ & $2k$ \\ 
& $6k + 3$ & ${\bf 2k+2}$ & $2k+1$ & $2k$ \\
& $6k+5$ & (${\bf 2k+1}$) & ${\bf 2k+2}$ & ${\bf 2k+2}$ \\
\hline 
\end{tabular} \hskip 2mm 
}
\label{tab:spectra}
\end{table}

A number theoretical comment is called for here, even though number theory will 
be kept in the background throughout this paper. We are interested in 
vectors whose components lie in a specified algebraic number field. Since we think 
of all our algebraic numbers as embedded in the complex field we can work in a 
projective space using the embedded number field as its field of scalars. The 
displacement operators and the symplectic group elements can be represented by 
unitary matrices all of whose matrix elements belong to the cyclotomic field, 
and this number field is a subfield of any number field that houses 
a SIC \cite{AFMY}. But consider a Zauner unitary $U_{\mathcal{Z}}$ of order 3 as an example. 
Then we get the eigenvectors by choosing 
any vector $|\psi\rangle$, and performing the projections 

\begin{equation} \begin{array}{lc} \frac{1}{3}({\bf 1} + U_{\mathcal{Z}} + 
U_{\mathcal{Z}}^2)|\psi \rangle & \\ \\ 
\frac{1}{3}({\bf 1} + \omega_3^2U_{\mathcal{Z}} + \omega_3U_{\mathcal{Z}}^2)|\psi \rangle 
& \\ \\ 
\frac{1}{3}({\bf 1} + \omega_3U_{\mathcal{Z}} + \omega_3^2U_{\mathcal{Z}}^2)|\psi \rangle & . 
\end{array} \end{equation}

\noindent Assume that the components of $|\psi\rangle$ lie in the specified number 
feld. If degenerate eigenvalues occur (as they will in this case) we can resort to 
Gram--Schmidt orthogonalization without leaving that number field. However, a 
complication arises for the symplectic unitaries because 
their eigenvalues may or may not belong to the 
desired field. In particular the third root of unity, which is an eigenvalue 
of the Zauner unitary, will belong to the relevant field in some cases (say if we start 
the ladder at $d_0 = 5$) but not in others (say if we start from $d_0 = 7$). 
This is a complication. Another complication arises if we normalize the eigenvectors, since 
the square root of the norm squared will typically not lie in the field. However, this 
does not mean that we cannot use normalized eigenvectors at intermediate stages in the 
construction of a SIC. If the ray class conjecture is true all the complications will 
cancel themselves out in the end. Examples can be found in ref.~\cite{SiSi}. 

We end this section with a few definitions from frame theory. A {\it frame} of type 
$(d,N)$ is a collection of $N$ vectors $\{ |u_I\rangle \}_{I = 1}^N$ that span 
${\bf C}^d$. The frame is {\it covariant} if the vectors form an orbit of a group. 
The frame is {\it tight} if there exists a real number $\alpha$ such that 

\begin{equation} \sum_{I=1}^N |u_I\rangle \langle u_I| = \alpha {\bf 1}_d \ . 
\end{equation}

\noindent The {\it generator matrix} of a frame is a $d\times N$ matrix $M_1$ 
whose columns are the vectors that make up the frame. Thus 

\begin{equation} M_1 = \left( \begin{array}{cccc} |u_1\rangle & |u_2\rangle & \dots & 
|u_N\rangle \end{array} \right) \ . \end{equation}

\noindent The condition that the 
frame be tight can then be written as $M_1M_1^\dagger = \alpha {\bf 1}$, which implies that 
the rows of the matrix be orthogonal to each other. The generator 
matrix can always be extended to a unitary matrix by adding $N-d$ additional rows, and 
by a suitable rescaling. Thus we have 

\begin{equation} U = \left( \begin{array}{c} M_1 \\ \hline M_2 \end{array} 
\right)_{N\times N} , \hspace{5mm} UU^\dagger = {\bf 1}_N \hspace{2mm} \Leftrightarrow 
\hspace{2mm} \left\{ \begin{array}{l} M_1M_1^\dagger = {\bf 1}_d \\ \\ 
\mbox{and} \\ \\ M_2M_2^\dagger = {\bf 1}_{N-d} \ . \end{array} \right. \end{equation}

\noindent We have then constructed a tight frame of type $(N-d,N)$ with generator matrix 
$M_2$, known as the {\it Naimark complement} of the frame we started out with. Note 
however that the complement is by no means uniquely defined. 
Insisting that a tight frame be equiangular leads to a difficult problem. But since 
the columns of a unitary matrix are orthonormal to each other the mutual overlaps 
among the vectors in the Naimark complement are determined by the mutual overlaps 
in the frame we started out with. In particular, if $M_1$ is the generator matrix 
of an ETF then so is $M_2$. 

\section{Aligned SICs and proto-SICs}\label{sec:aligned}

\

\noindent We now slow down a little since the ladders themselves do deserve a more 
detailed introduction. The number theoretical 
reasoning that suggests the existence of the dimension ladders begins with the observation 
that the number field needed to construct a SIC in dimension $d$ is an extension of the 
number field ${\bf Q}(\sqrt{D})$, where $D$ is the square free part of the integer 
$(d+1)(d-3)$ \cite{AYAZ}. But it is easy to see that 

\begin{equation} (d(d-2)+1)(d(d-2)-3) = (d-1)^2(d+1)(d-3) \ . \end{equation}

\noindent It follows that the square free part of $(d+1)(d-3)$ is unchanged 
under the substitution $d \rightarrow d(d-2)$. The ray class conjecture 
then implies that the minimal number field needed to construct a SIC in dimension 
$d$ is a subfield of the minimal number field needed to construct a SIC in 
dimension $d(d-2)$, essentially because the conductor $d$ divides 
the conductor $d(d-2)$. To avoid misunderstanding we should add 
that the sequences $d_{n+1} = d_n(d_n-2)$ are only subsequences of the towers of 
dimensions that are related to a given real quadratic field \cite{AFMY}. 

If $d$ is odd then $d$ and $d-2$ are relatively prime and the Weyl--Heisenberg 
group splits as a direct product. We focus on the displacement operators, and find 
following equation (\ref{Kina}) that 

\begin{equation} D^{d(d-2)}_{\bf p} = D^{(d-2)}_{{\bf p}'}\otimes D^{(d)}_{{\bf p}''} \ 
. \end{equation}

\noindent The symplectic unitaries behave similarly. Hence we can write 

\begin{equation} {\bf C}^{d(d-2)} = {\bf C}^{d-2}\otimes {\bf C}^d \end{equation}

\noindent in a meaningful way. In even dimensions complications arise at this point. 
With appropriate measures taken the basic logic is unchanged \cite{AD}, but these 
measures are subtle and we have decided to postpone a discussion of even 
dimensions to a later occasion. 

From now on we work in ${\bf C}^{d-2}\otimes {\bf C}^d$ with $d$ odd. Paraphrasing the 
definition in a previous paper \cite{aligned} we define aligned SICs as follows:

\

\noindent \underline{Definition 1}: {\sl A SIC with fiducial vector $|\Psi_{\bf 0}\rangle$ 
in dimension $d(d-2)$ is {\it aligned} to a SIC with overlap phases 
$e^{i\theta_{\bf p}}$ in an odd dimension $d$ if} 

\begin{equation} (d-1)\langle \Psi_0|{\bf 1}_{d-2}\otimes D^{(d)}_{\bf p}|\Psi_0 \rangle 
= - e^{2i\theta_{M{\bf p}}} \label{aligned1} \end{equation}

\noindent {\sl and}

\begin{equation} (d-1)\langle \Psi_0|D^{(d-2)}_{\bf p}\otimes {\bf 1}_d|\Psi_0 \rangle 
= 1 \  , \label{aligned2} \end{equation}

\noindent {\sl for a matrix $M$ with determinant $\pm 1$ modulo $d$.} 

\

Recall that $d-1 = \sqrt{d(d-2)+1}$. As we will show below, the key condition is 
equation (\ref{aligned1}). The other follows as a corollary. The precise linear 
relation between ${\bf p}$ and $M{\bf p}$ is slightly complicated by the symplectic automorphism group, 
which is why $M$ is not completely fixed by the definition. 
The available evidence \cite{aligned} suggests that {\it every} SIC in dimension 
$d$ has an aligned counterpart in dimension $d(d -~2)$. 

We make one more definition: 

\

\noindent \underline{Definition 2}: {\sl A proto-SIC in dimension $d(d-2)$ is a 
Weyl--Heisenberg orbit of a vector $|\Psi_0\rangle$ obeying equations 
(\ref{aligned1}) and (\ref{aligned2}).}

\

An aligned SIC is a proto-SIC, but the converse need not hold. In this paper 
we will provide a recipe for how to calculate proto-SICs, given a SIC in dimension 
$d$ to start from. In fact we will obtain a continuous family of proto-SICs. We will 
{\it not} provide a proof that the parameters can be chosen so that the SIC condition 

\begin{equation} (d-1)^2|\langle \Psi_0|D^{(d-2)}_{{\bf p}^\prime}\otimes 
D^{(d)}_{{\bf p}{\prime \prime}}|\Psi_0 \rangle |^2 = 1 \end{equation}

\noindent holds as well. 

Still it is interesting to pause and consider the geometric meaning of proto-SICs. It 
is known \cite{aligned} that the $d^2$ vectors obtained by acting with ${\bf 1}_{d-2} 
\otimes D_{\bf p}^{(d)}$ on $|\Psi_0\rangle$ span a subspace of dimension 
$d(d-1)/2$ in ${\bf C}^{d(d-2)}$. Hence they form an equiangular tight frame in this 
subspace. As a consistency check on the proof we examine equation 
(\ref{etfs}). By replacing $d \rightarrow d(d-1)/2$, and setting $N = d^2$, we obtain 

\begin{equation} |\langle \psi_I|\psi_J\rangle|^2 = \frac{d^2-\frac{d(d-1)}{2}}
{\frac{d(d-1)}{2}(d^2-1)} = \frac{1}{(d-1)^2} = \frac{1}{d(d-2)+1} \ . \end{equation}

\noindent The right hand side is just right for a pair of SIC vectors in dimension 
$d(d-2)$. Acting on the vectors in this ETF with group elements of the form 
$D^{(d-2)}_{\bf p}\otimes {\bf 1}_d$ we obtain a set of $(d-2)^2$ equiangular 
tight frames, all of them spanning some $d(d-1)/2$ dimensional subspace. We can 
regard the resulting set of subspaces as so many points in a Grassmannian, 
and then it can be shown that these points are equidistant with respect to 
the natural metric on the Grassmannian \cite{explanations}. 

An aligned SIC therefore has such a structure sitting inside it. There is an 
alternative partitioning of the aligned SIC into $d^2$ sets 
of $(d-2)^2$ vectors forming ETFs in subspaces of dimension $(d-2)(d-1)/2$, but this 
will not play a role in the arguments to follow. 

Aligned SICs have found some use. For instance, the exact solution in dimension 
$323 = 19\cdot 17$ was found using the guidance of equations (\ref{aligned1}) 
and (\ref{aligned2}) \cite{GS}. But we want more---more, in fact, than we can provide 
at the moment. In Sections \ref{sec:theorem1} and \ref{sec:theorem2}  
we will present a chain of theorems that allows us to calculate a fiducial vector 
$|\Psi_0\rangle$ obeying equations (\ref{aligned1}) and (\ref{aligned2}), given any 
SIC in dimension $d$ to start from. The key to the construction is the 
calculation of the equiangular tight frames that are to be found embedded in the
higher dimensional SIC. Unfortunately no chain is stronger than its weakest link. 
The weak link in our chain will turn out to be that the construction contains 
ambiguities and leads to a continuous family of fiducial vectors. We repeat that 
we have not proven that there is a SIC present in every such family of proto-SICs. 

\section{General theorems}\label{sec:theorem1}

\

\noindent In the discovery papers it was proved that a SIC is a complex projective 2--design 
\cite{Zauner, Renes}.  We can restate this as a theorem that starts our chain: 

\

%

\begin{theorem}\label{Theorem1}
{\sl It holds that} 

\begin{equation} \{ |\psi_I\rangle \}_{I=1}^{d^2} \ \mbox{is a SIC} \hspace{2mm} 
\Leftrightarrow \hspace{2mm} 
\{ |\psi_I\rangle \otimes |\psi_I\rangle \}_{I=1}^{d^2} \ \mbox{a tight frame in} \ 
{\bf C}^d\otimes_S {\bf C}^d . 
\end{equation}
\end{theorem}

\noindent \underline{Proof}: This is a restatement of the 2--design property. We are using the 
formulation from section 2 in ref. \cite{Kiev}.  $\Box$

\

If the $d$-dimensional SIC is an orbit under the Weyl--Heisenberg group so is the 
ETF in the symmetric subspace, although in the latter case the representation is by 
necessity reducible. We will need a definite prescription for it. To find one we 
introduce the generators of the group, and observe that 

\begin{equation} (Z\otimes Z)(X\otimes X) = \omega^2 (X\otimes X)(Z\otimes Z) \ . 
\label{xxzz} \end{equation}

\noindent There is a clear dichotomy between odd and even dimensions. In odd 
dimensions $\omega_d^2$ remains a $d$th root of unity, and the symmetric product group 
is isomorphic to the Weyl--Heisenberg group we start out with. This is not so in 
even dimensions, but we have already made the restriction to odd $d$. We can then 
define the operators 

\begin{equation} {\bf X} = X\otimes X \hspace{5mm} \mbox{and} \hspace{5mm} 
{\bf Z} = Z^\frac{d+1}{2} \otimes Z^\frac{d+1}{2} \end{equation}

\noindent to generate the group when acting on ${\bf C}^d\otimes {\bf C}^d$. Then 
there holds 

\begin{equation} {\bf Z}{\bf X} = \omega {\bf X}{\bf Z} \ . \end{equation}

\noindent We define the displacement operators 

\begin{equation} \tilde{D}_{i,j} = \tau^{ij}{\bf X}^i{\bf Z}^j = 
D_{i,2^{-1}j}\otimes D_{i,2^{-1}j} \ . \end{equation}

\noindent Recall that the arithmetic of the labelling integers is modulo $d$. 
It will be convenient to introduce the matrix 

\begin{equation} H = \left( \begin{array}{cc} 1 & 0 \\ 0 & 2^{-1} \end{array} \right) \ , 
\label{H} \end{equation}

\noindent and write this as 

\begin{equation} \tilde{D}_{\bf p} = D_{H{\bf p}}\otimes D_{H{\bf p}} \ . \end{equation}

We will decompose this reducible representation as a direct sum of $d$-dimensional 
Weyl representations. First we introduce the basis vectors 

\begin{eqnarray} & |(i,i)\rangle = |e_i\rangle \otimes |e_i\rangle \hspace{31mm} \\ 
\nonumber \\ 
& |(i,j)\rangle = \frac{1}{\sqrt{2}}(|e_i\rangle \otimes |e_j\rangle + 
|e_j\rangle \otimes |e_i\rangle ) \ \\ \nonumber \\  
& |[i,j]\rangle = \frac{1}{\sqrt{2}}(|e_i\rangle \otimes |e_j\rangle - 
|e_j\rangle \otimes |e_i\rangle ) \ .  \end{eqnarray}

\noindent Then we can decompose ${\bf C}^{d}\otimes {\bf C}^d$ into the symmetric 
and anti-symmetric subspaces at our convenience. If the standard representation of 
the displacement operators is being used we can take the representation to be carried 
by the $d$ subspaces spanned by 

\begin{eqnarray} |( x\frac{d-1}{2}+i,x\frac{d+1}{2}
+ i ) \rangle \ , \hspace{8mm} 0 \leq x \leq \frac{d-1}{2} \ \ 
\label{symbas} \\ \nonumber \\ |[ x\frac{d-1}{2}+i,x\frac{d+1}{2}
+ i ]\rangle \ , \ \hspace{8mm} 1 \leq x \leq \frac{d-1}{2} \ . \end{eqnarray} 

\noindent To check that this works it is enough to check that the action of the 
generators ${\bf X}$ and ${\bf Z}$ works as expected, in particular that they 
leave the label $x$ invariant. 

At this point we may wish to forget about the tensor product structure we 
started out with. Instead we replace it with a new one, so that the representation can 
be written 

\begin{equation} \tilde{D}_{\bf p} = {\bf 1}_d\otimes D_{\bf p} \ . \label{odddisp} \end{equation}

\noindent In this way the group provides a natural decomposition of ${\bf C}^{d^2}$ as a 
direct sum of $d$ copies of ${\bf C}^d$. We can restrict ourselves to the symmetric 
subspace by letting the unit operator act on a space of dimension $(d+1)/2$. 

We are interested in covariant frames made from vectors of the form 
$\tilde{D}_{\bf p}{\bf u}$, where the fiducial vector ${\bf u}$ is a direct sum of 
$k$ vectors ${\bf x}_r \in {\bf C}^d$. This is to say that the generator matrices 
are $dk\times d^2$ matrices formed according to 

\begin{equation} M_1 = \left( 
\begin{array}{cccc} {\bf x}_0 & D_{0,1}{\bf x}_0 & \dots & D_{d-1,d-1}{\bf x}_0 \\ 
\vdots & \vdots & & \vdots \\ 
{\bf x}_{k-1} & D_{0,1}{\bf x}_{k-1} & \dots & D_{d-1,d-1}{\bf x}_{k-1} 
\end{array} \right)_{dk\times d^2} \ . \label{M1} \end{equation}

\noindent Recall from Section 2 that the columns of $M_1$ form  a tight frame in 
${\bf C}^{kd}$ if and only if $M_1M_1^\dagger = d{\bf 1}_{kd}$. 

%

\begin{theorem}\label{Theorem2}  
{\sl The frame with generator matrix $M_1$ is tight if 
and only if the vectors ${\bf x}_i$ form an orthonormal set in ${\bf C}^d$}, 

\begin{equation} M_1M_1^\dagger = d{\bf 1}_{kd} \hspace{5mm} \Leftrightarrow \hspace{5mm} 
\langle {\bf x}_s, {\bf x}_r\rangle = \delta_{rs} \ . \end{equation}
\end{theorem}

\

\noindent \underline{Proof}: For any unitary operator basis $\{ U_I\}_{I = 1}^{d^2}$, for 
any operator $A$, and for any pair of vectors ${\bf x}_i, {\bf x}_j$ there holds 

\begin{equation} \sum_I U_IAU_I^\dagger = d\mbox{Tr}A{\bf 1}_d \hspace{5mm} \mbox{and} 
\hspace{5mm} \sum_I U_I{\bf x}_r{\bf x}_s^\dagger U_I^\dagger = 
\langle {\bf x}_s, {\bf x}_r\rangle d{\bf 1}_d \ . \label{UOB} \end{equation}

\noindent Because the displacement operators form a unitary operator basis the proof 
of the theorem follows by inspection. $\Box$

\

Following Section 2 we can find the Naimark complement by adding an additional set of 
$d(d-k)$ rows to $M_1$, obtaining a matrix that is unitary up to an overall factor. 
Explicitly 

\begin{equation} U = \left( \begin{array}{c} M_1 \\ \hline M_2\end{array} \right) = 
\left( \begin{array}{cccc} {\bf x}_0 & D_{0,1}{\bf x}_0 & \dots & D_{d-1,d-1}{\bf x}_0 \\ 
\vdots & \vdots & & \vdots \\ 
{\bf x}_{k-1} & D_{0,1}{\bf x}_{k-1} & \dots & D_{d-1,d-1}{\bf x}_{k-1} \\ \hline 
{\bf x}_{k} & D_{0,1}{\bf x}_{k} & \dots & D_{d-1,d-1}{\bf x}_{k} \\
\vdots & \vdots & & \vdots \\ 
{\bf x}_{d-1} & D_{0,1}{\bf x}_{d-1} & \dots & D_{d-1,d-1}{\bf x}_{d-1}
\end{array} \right)_{d^2\times d^2} \ . \label{M2} \end{equation} 

\noindent The frame whose generator matrix is $M_2$ will be
tight provided only that the vectors $\{ {\bf x}_r\}_{r=k}^{d-1}$ fill out an orthonormal 
basis $\{{\bf x}_r\}_{r=0}^{d-1}$ in ${\bf C}^d$. This can be arranged with Gram--Schmidt, 
if no better alternative offers itself. 

\

\section{General theorems, continued}\label{sec:theorem2}

\

\noindent We begin a new section here because from now on we will specialize the 
vectors ${\bf x}_r$ from which our tight frames are constructed. 

In Theorem \ref{Theorem1} we obtained not only a tight frame but a group covariant equiangular 
tight frame in the symmetric subspace of ${\bf C}^d\otimes {\bf C}^d$. Its fiducial 
vector is

\begin{equation} |\tilde{\Psi}\rangle = |\psi_0\rangle \otimes |\psi_0 \rangle \ , 
\end{equation}

\noindent where $|\psi_0\rangle$ is the fiducial vector of a SIC in dimension $d$. Its overlaps are 

\begin{equation} \langle \tilde{\Psi}|\tilde{D}_{\bf p}|\tilde{\Psi}\rangle = 
\left( \langle \psi_0|D_{H{\bf p}}|\psi_0\rangle \right)^2 = 
\frac{e^{2i\theta_{H {\bf p}}}}{d+1} \ . \label{overlap1} \end{equation} 

\noindent Using the basis (\ref{symbas}) and rewriting $\tilde{D}_{\bf p}$ as 
${\bf 1}\otimes D_{\bf p}$ we obtain an ETF of the form presented in Theorem 
\ref{Theorem2}, with $k = (d+1)/2$. Hence, from now on, 

\begin{equation} \left( \begin{array}{c} {\bf x}_0 \\ {\bf x}_1 \\ \vdots \\ 
{\bf x}_{k-1} \end{array} \right) = \sqrt{\frac{d+1}{2}} |\psi_0\rangle \otimes |\psi_0 
\rangle \ , \hspace{8mm} k = \frac{d+1}{2} \ , \end{equation}

\noindent where it is understood that $|\psi_0\rangle \otimes |\psi_0 \rangle$ has been 
expressed in the basis (\ref{symbas}). Hence the components of the $(d+1)/2$ vectors 
$\{ {\bf x}_r\}_{r=0}^{k-1}$ are known quadratic functions of the components of 
the SIC fiducial vector $|\psi_0\rangle$. 

The overlaps are not of the right size for this ETF to be embedded in a $d(d-2)$ 
dimensional SIC. This is easily remedied by taking the Naimark complement. This gives 
us an ETF in dimension $d(d-1)/2$, with fiducial vector 

\begin{equation} |\Psi_{\bf 0} \rangle = \sqrt{\frac{2}{d-1}} \left( \begin{array}{c} 
{\bf x}_k \\ {\bf x}_{k+1} \\ \vdots \\ {\bf x}_{d-1} \end{array} \right) \ , 
\hspace{8mm} k = \frac{d+1}{2} \ . \end{equation}

\noindent By construction the columns of the matrix $U$ defined in equation (\ref{M2}) 
are orthogonal. Given equation (\ref{overlap1}), and keeping track of the normalizing 
factors, we conclude that 

\begin{equation} \langle \Psi_{\bf 0}|\tilde{D}_{\bf p}|\Psi_{\bf 0}\rangle = - 
\frac{e^{2i\theta_{H {\bf p}}}}{d-1} \ . \label{kvadoverlapp} \end{equation} 

\noindent We have arrived at 

%

\begin{theorem}\label{Theorem3}
{\sl Given a Weyl--Heisenberg covariant SIC in 
dimension $d$, with overlap phases $e^{i\theta_{\bf p}}$, an ETF with $d^2$ 
vectors in dimension $d(d-1)/2$ and overlap phases given by equation 
(\ref{kvadoverlapp}) can be obtained by direct calculation.}
\end{theorem}

\

\noindent \underline{Proof}: Given already. It only remains to add that the argument was 
inspired by Ostrovskyi and Yakymenko \cite{Kiev}. $\Box$ 

\

The construction of the Naimark complement is not unique. We have the freedom to choose 
an orthonormal basis in a subspace of dimension $d(d-1)/2$. Ignoring an overall phase 
this leaves us with a family of solutions parametrized by the group $SU((d-1)/2)$. 
Theorem 5 below will provide a nice interpretation of this freedom. 

We begin by collecting 
all the available information about the subspace of ${\bf C}^d$ where the basis is to be 
chosen. Also, following the best of examples \cite{Dirac}, we switch to bra-ket 
notation for all vectors.

%

\begin{theorem}\label{Theorem4}
{\sl There holds} 

\begin{equation} \sum_{r=0}^{\frac{d-1}{2}}|x_r\rangle \langle x_r| = \frac{1}{2} 
\left( {\bf 1} + P_\theta \right) \label{Ptheta1} \end{equation}

\noindent {\sl where $P_\theta$ obeys $P^2_\theta = {\bf 1}$ and is given by}

\begin{equation} P_\theta = \frac{1}{d}\sum_{\bf p}e^{2i\theta_{H{\bf p}}}D_{-{\bf p}} \ . 
\label{Ptheta2} \end{equation}
\end{theorem}

\

\noindent \underline{Proof}: Expand the projection operator as 

\begin{equation} \sum_{r=0}^{\frac{d-1}{2}}|x_r\rangle \langle x_r| = 
\frac{1}{d}\sum_{\bf p}\sum_r\langle x_r|D_{\bf p}|x_r\rangle D_{-{\bf p}} \ . \end{equation}

\noindent The squared SIC phases enter when we use the fact that 

\begin{equation} \sum_{r=0}^{\frac{d-1}{2}}\langle x_r|D_{\bf p}|x_r\rangle = \frac{d+1}{2} 
\langle \psi_0|\langle \psi_0|\tilde{D}_{\bf p}|\psi_0\rangle |\psi_0\rangle \ . 
\end{equation} 

\noindent Full details are given by Ostrovskyi and Yakymenko, in their Section 7 
\cite{Kiev}. $\Box$

\

The operator $P_{\theta}$ is known as the generalised parity operator \cite{aligned}. It 
played a significant role in ref. \cite{Dardo}, where it was used to construct ETFs at 
the Gram matrix level. Equation (\ref{Ptheta1}) provides a convenient way to calculate 
it from the components of a $d$ dimensional SIC vector. 

The important point is that we have divided ${\bf C}^d$ into a direct sum, 

\begin{equation} {\bf C}^d = {\bf C}^\frac{d+1}{2}\oplus {\bf C}^\frac{d-1}{2} \equiv 
\mathcal{H}_{P_\theta}^{(+)} \oplus \mathcal{H}_{P_\theta}^{(-)} \end{equation}

\noindent (where we took the opportunity of introducing some new notation), and that this 
decomposition into eigenspaces carries information about the SIC overlap phases from 
dimension $d$. 

In dimension $d-2$ there is a similar decomposition into 

\begin{equation} {\bf C}^{d-2} = {\bf C}^\frac{d-1}{2}\oplus {\bf C}^\frac{d-3}{2} \equiv 
\mathcal{H}_{P}^{(+)} \oplus \mathcal{H}_{P}^{(-)} \end{equation}

\noindent defined by the standard parity operator $U_P$. See Table \ref{tab:spectra}. 

We now have a family of ETFs with overlaps of the desired size. Moreover they carry some 
information about the SIC in dimension $d$. The orthonormal vectors from which their fiducial 
vector is built span $\mathcal{H}_{P_\theta}^{(-)} \in {\bf C}^d$, an eigenspace of 
the generalised parity operator $P_\theta$. The next step is to embed the fiducial 
vector in the Hilbert space ${\bf C}^{d-2}\otimes {\bf C}^d$. 

For convenience we rename the basis vectors in $\mathcal{H}^{(-)}_{P_\theta}$ according to 

\begin{equation} {\bf x}_{\frac{d+1}{2}} \ \rightarrow \ |e_0\rangle \ , \ \dots \ , 
\ {\bf x}_{d-1} \ \rightarrow \ |e_{\frac{d-1}{2}-1}\rangle \ . \label{adapt} \end{equation}

\noindent When we introduce orthonormal basis vectors in $\mathcal{H}_{P}^{(+)}$ (which has 
the same dimension as $\mathcal{H}_{P_\theta}^{(-)}$) we will denote them by $|f_r\rangle $. 
Should we wish to climb more than one rung of the ladder, and work in a Hilbert space 
built from more than two factors, we will introduce further basis vectors 
$|g_r\rangle$, $|h_r\rangle$, and so on. We do not believe that explicit calculations 
for more than three rungs will be performed any time soon, and hence we foresee 
no notational problems in the near future. 

Now consider the tensor product Hilbert space 

\begin{equation} \mathcal{H}_{P}^{( +)}\otimes \mathcal{H}_{P_\theta}^{( -)} \in 
{\bf C}^{d-2}\otimes {\bf C}^d \ . \end{equation}

\noindent In section 5 of ref. \cite{aligned} it was proved that every aligned SIC has a fiducial 
vector that is a maximally entangled vector in this subspace. Here we are more interested in 
the converse:  

%

\begin{theorem}\label{Theorem5}
{\sl Every maximally entangled unit vector $|\Psi_{\bf 0}\rangle 
\in \mathcal{H}_{P}^{( +)} \otimes \mathcal{H}_{P_\theta}^{( -)}$ obeys}  

\begin{equation} (d-1)\langle \Psi_{\bf 0}|{\bf 1}_{d-2}\otimes D^{(d)}_{\bf p}|\Psi_{\bf 0} \rangle 
= - e^{2i\theta_{{\bf p}'}} \label{teorema} \end{equation} 

\begin{equation} (d-1)\langle \Psi_{\bf 0}|D^{(d-2)}_{\bf p}\otimes {\bf 1}_d|\Psi _{\bf 0}\rangle 
= 1 \label{teoremb} \end{equation}

\noindent {\sl for ${\bf p} \neq {\bf 0}$, where ${\bf p}'$ is linearly related to ${\bf p}$.} 
\end{theorem}

\

\noindent \underline{Proof}: The proof uses the Schmidt decomposition of the state. 
We introduce the adapted basis (\ref{adapt}) and write the state on the form 

\begin{equation} |\Psi_{\bf 0}\rangle = \sqrt{\frac{2}{d-1}}\ \sum_{r=0}^{(d-1)/2-1} 
|f_r\rangle \otimes |e_r\rangle \ , \end{equation}

\noindent By definition of `maximally entangled' the basis vectors $|f_r\rangle$ can 
always be found. We find 

\begin{equation} (d-1) \langle \Psi_{\bf 0}|{\bf 1}_{d-2}\otimes D^{(d)}_{\bf p}|\Psi_{\bf 0} \rangle = 
\sum_r\langle e_r|D_{\bf p}^{(d)}|e_r\rangle = - e^{2i\theta_{{\bf p}^\prime}} \end{equation}

\noindent by preceding results. We also find 

\begin{eqnarray} (d-1) \langle \Psi_{\bf 0}|D^{(d-2)}_{\bf p}\otimes {\bf 1}_d
|\Psi_{\bf 0} \rangle = \sum_r\langle f_r|D_{\bf p}^{(d-2)}|f_r\rangle = \nonumber \\ \\ 
= \frac{1}{2}\mbox{Tr}D_{\bf p}^{(d-2)}({\bf 1}_{d-2} + U_P) = 1 \ . \nonumber \end{eqnarray}

\noindent This holds because the basis vectors are positive parity eigenstates, 
and because Tr$D_{\bf p}U_P = 1$ for all ${\bf p}$ in all odd dimensions. $\Box$

\

It is well known that the set of maximally entangled states in a Hilbert space 
of the form ${\bf C}^N\otimes {\bf C}^N$ can be obtained by choosing orthonormal bases 
in the two factors, and writing the states in the form 

\begin{equation} |\Phi\rangle = \frac{1}{\sqrt{N}}\sum_{r,s}U_{r,s}|r\rangle \otimes 
|s\rangle \label{snarjelse} \end{equation}

\noindent where $U$ is any unitary matrix. Hence the set of maximally entangled states is 
isomorphic to the group manifold $SU(N)/Z_N$, where an overall phase factor was removed 
by factoring out the discrete subgroup $Z_N$. By adapting the bases to the state the matrix 
$U$ can be written in diagonal form, which is how the Schmidt form of the state arises. 
In this way Theorem \ref{Theorem5} provides us with an 
alternative view of the freedom observed in Theorem \ref{Theorem3}. 

We are now in a position to calculate a set of vectors in dimension $d(d-2)$ that obey 
equations (\ref{aligned1}) and (\ref{aligned2}), given a SIC in dimension $d$ to start 
with. But the freedom we have in doing this is large while the number of SICs is 
expected to be finite. To cut down on the ambiguities we will appeal 
to the symmetries of the $d$-dimensional SIC. This symmetry survives (and is in fact 
enhanced) when we go to the next rung of the ladder \cite{aligned, AD}. Here we will 
rely on:

%
\begin{theorem}\label{Theorem6}
{\sl For a symplectic unitary $U_F$ it holds that}

\begin{equation} U_F|\psi_0\rangle \sim |\psi_0\rangle \hspace{5mm} \Rightarrow 
\hspace{5mm} \left[ U_{F^\prime}, P_\theta \right] = 0 \ , \end{equation}

\noindent {\sl where} $\sim$ {\sl means equal up to a phase factor, and} $F^\prime = 
H^{-1}FH$ {\sl where} $H$ {\sl is the matrix defined in equation (\ref{H}).} 
\end{theorem}

\

\noindent \underline{Proof}: Suppose that $|\psi_0\rangle$ is an eigenvector of 
$U_F$. We then find that 

\begin{eqnarray} \langle \psi_0|\langle \psi_0|\tilde{D}_{\bf p}|\psi_0\rangle 
|\psi_0\rangle = \langle \psi_0|\langle \psi_0|U_FD_{H{\bf p}}U_F^{-1}\otimes 
U_FD_{H{\bf p}}U_F^{-1}|\psi_0\rangle |\psi_0\rangle = \nonumber \\ \nonumber \\ 
= \langle \psi_0|\langle \psi_0|D_{FH{\bf p}}\otimes 
D_{FH{\bf p}}|\psi_0\rangle |\psi_0\rangle = \hspace{8mm} \\ \nonumber \\ = 
\langle \psi_0|\langle \psi_0|D_{HH^{-1}FH{\bf p}}\otimes 
D_{HH^{-1}FH{\bf p}}|\psi_0\rangle |\psi_0\rangle \ . 
\nonumber \end{eqnarray}

\noindent We define 

\begin{equation} F^\prime = H^{-1}FH \ , \end{equation}

\noindent and rewrite the equation (using the second tensor product structure) as 

\begin{equation} \langle u|{\bf 1}\otimes D_{\bf p}|u\rangle = \langle u|{\bf 1} 
\otimes D_{F^\prime {\bf p}}|u\rangle = \langle u|{\bf 1}\otimes U_{F^\prime}
D_{\bf p}U_{F^\prime}^{-1}|u\rangle \ . \end{equation}

\noindent Given this, we can proceed to show that 

\begin{eqnarray} U_{F^\prime}^{-1}\frac{1}{2}({\bf 1} + P_\theta )U_{F^\prime} = 
U_{F^\prime}^{-1}\sum_r|x_r\rangle \langle x_r|U_{F^\prime} = \hspace{30mm} \nonumber \\ 
\nonumber \\ 
= \frac{1}{d}\sum_{\bf p}\sum_r\langle x_r|U_{F^\prime}D_{\bf p}U^{-1}_{F^\prime}
|x_r\rangle D_{-{\bf p}} = \frac{1}{d}\sum_{\bf p}\langle u|{\bf 1}\otimes U_{F^\prime}
D_{\bf p}U^{-1}_{F^\prime}|u\rangle D_{-{\bf p}} = \\ \nonumber \\ 
= \frac{1}{d}\sum_{\bf p}\langle u|{\bf 1}\otimes D_{\bf p}|u\rangle D_{-{\bf p}} 
= \frac{1}{d}\sum_{\bf p}\sum_r\langle x_r|D_{\bf p}|x_r\rangle D_{-{\bf p}} = 
\frac{1}{2}({\bf 1} + P_\theta ) \ . \nonumber 
\end{eqnarray}

\noindent Hence $U_{F^\prime}$ commutes with $P_\theta$. $\Box$

\

The ambiguities in the construction of the ETF will be cut down considerably 
if we insist that the vectors $|e_r\rangle $ from which it is built are eigenvectors of 
$U_{F'}$. In the Schmidt form of the fiducial vector to be built in ${\bf C}^{d-2}\otimes 
{\bf C}^d$ we are free to use basis vectors in $\mathcal{H}_P^{(+)}$ in such a way that 
the fiducial vector becomes an eigenvector of a symplectic unitary $U_{F''}\otimes U_{F'}$. 
In this way we can arrange that its symmetry group has the same order as that of the 
SIC fiducial in ${\bf C}^d$. In fact its order will be twice that because by construction 
the fiducial is invariant under $U_P\otimes {\bf 1}_d$. (This is the $F_b$ symmetry 
in the notation of Scott and Grassl \cite{Scott}.) The upshot is that the construction 
will use quite special maximal entangled states for which the unitary matrix in equation (\ref{snarjelse}) becomes block-diagonal. 

\

\section{Instructions for how to climb a ladder}\label{sec:recipe}

\

\noindent We are ready to climb the ladders. Regardless of the somewhat lengthy argument 
that we have gone through the procedure is quite simple, so we spell it out here:

\

\noindent {\bf 1}. Choose a SIC in dimension $d$ invariant under a symplectic 
unitary $U_{F}^{(d)}$. Compute the generalised parity operator $P_\theta$ as 
well as the Zauner unitary $U^{(d)}_{F^\prime}$ defined 
in Theorem \ref{Theorem6}. 

\

\noindent {\bf 2}. Use the parity operators $U_P$ and $P_\theta$ to define the positive 
parity subspace of ${\bf C}^{d-2}$ and the negative generalised parity subspace of 
${\bf C}^d$. Their dimensions are equal to $(d-1)/2$ in both cases. 

\

\noindent {\bf 3}. Compute bases for the subspaces defined in the previous step, as 
eigenvectors of $({\bf 1} + U_P)U^{(d-2)}_{F^{\prime \prime}}$ and of 
$({\bf 1} - P_\theta )U^{(d)}_{F^\prime}$, where $U^{(d-2)}_{F^{\prime \prime}}$ 
is some chosen Zauner unitary having the same order as $U_{F^\prime}^{(d)}$. 

\

\noindent {\bf 4}. Use these bases to form maximally entangled states in 
${\bf C}^{(d-1)/2}\otimes~{\bf C}^{(d-1)/2}$ considered as a subspace of 
${\bf C}^{d-2}\otimes {\bf C}^d$, and arrange them so that they lie in an eigenspace 
of $U_{F^{\prime \prime}}^{(d-2)}\otimes U_{{F}^\prime}^{(d)}$. 

\

\noindent {\bf 5}. The resulting family of proto-SICs is parametrized by a block 
diagonal unitary matrix. Search for a SIC within this family. 

\

\noindent There are a few useful hints that should be digested before one 
applies this recipe. Rather than spelling them all out at once we turn to examples. 

\

\section{A simple example}\label{sec:example}

\

\noindent We first choose $d_0=5$. There is a unique Clifford orbit of SICs 
labelled 5a \cite{Scott}, having a symmetry of order 3. Each eigenspace 
defined by the standard Zauner matrix with eigenvalues chosen 
according to Table \ref{tab:spectra} contains four distinct but Clifford 
equivalent SIC fiducial vectors. We choose the one given in equation (4) in ref. 
\cite{SiSi}. It is invariant under the standard Zauner unitary corresponding to 
the Zauner matrix (\ref{Zaunermatris}), and it determines the $(d+1)/2$ vectors $|x_r\rangle$ 
appearing in equation (\ref{Ptheta1}). We use this equation to 
calculate the generalised parity operator $P_{\theta}$. We also calculate 
the Zauner unitary $U_{\mathcal{Z}^\prime}^{(5)}$ appearing in Theorem \ref{Theorem6}. 
We then calculate the eigenvectors of $U_{\mathcal{Z}^\prime}^{(5)}
({\bf 1} -P_\theta )$. When we have discarded the ones that correspond to eigenvalue 
zero there are two such vectors left, and we label them with the eigenvalues of 
$U_{\mathcal{Z}^\prime}^{(5)}$. That is 

\begin{equation} \begin{array}{l} |e_0\rangle = |\omega_3\rangle_5 \\ 
|e_1\rangle = |\omega_3^2\rangle_5 \ . \end{array} \end{equation}

\noindent Since this step follows a standard procedure we do not give the details. 
It is however of importance to ensure that no spurious phase factors arise here, 
especially if one intends to convert a numerical solution to an exact one at the 
end. For this reason we insist that the first components of all the basis vectors 
we introduce are real. (Should the first component of some eigenvector vanish we 
set the second components of all the vectors in that basis to be real.) 

We then factor in the Hilbert space ${\bf C}^3$, and choose a Zauner unitary 
there. If possible we choose one that is represented by a monomial matrix. An 
appropriate choice is \cite{SiSi}

\begin{equation} \mathcal{Z}^{\prime \prime} = 
\left( \begin{array}{rr} 1 & 0 \\ 1 & 1 \end{array}\right)_3 \hspace{5mm} 
\Rightarrow \hspace{5mm} U_{\mathcal{Z}}^{(3)} = \left( \begin{array}{ccc} \omega_3^2 & 0 & 0 \\ 
0 & 1 & 0 \\ 0 & 0 & 1 \end{array} \right) \ . \end{equation}

\noindent There is a further choice made here, since we could replace 
$\mathcal{Z}^{\prime \prime}$ with its square. We have tried both possibilities 
and we have chosen the one that works. We calculate the eigenvectors of 
$U_{\mathcal{Z}^{\prime \prime}}^{(3)}({\bf 1} + U_P)$ and again label the eigenvectors 
by their eigenvalues, 

\begin{equation} \begin{array}{l} |f_0\rangle = |1\rangle_3 \\ |f_1\rangle = 
|\omega_3^2\rangle_3 \ . \end{array} \end{equation}

\noindent From these basis vectors we can construct maximally entangled states in 
${\bf C}^2\otimes {\bf C}^2 \subset {\bf C}^3\otimes {\bf C}^5$, and hence families of 
proto-SICs. 

We want a family of proto-SICs that contains a SIC. For this purpose we choose 
the maximally entangled states so that they belong to an eigenspace of the 15 dimensional 
Zauner unitary $U_{\mathcal{Z}^{\prime \prime}}\otimes 
U_{\mathcal{Z}^\prime}$ in ${\bf C}^3\otimes {\bf C}^5$. Hence we choose 

\begin{equation} |\Psi_{\bf 0} (\sigma )\rangle = \frac{1}{\sqrt{2}}\left( |f_0\rangle |e_0\rangle 
+ e^{i\sigma }|f_1\rangle |e_1\rangle \right) \ . \label{proto15} \end{equation}

\noindent The question now arises if this family contains a SIC. Thus we are looking 
for values of $\sigma$ such that 

\begin{equation} \sum_{{\bf p} \neq {\bf 0}} \left( |\langle \Psi_{\bf 0} (\sigma ) 
|D^{(d)}_{\bf p}|\Psi_{\bf 0} (\sigma) \rangle |^2 - \frac{1}{d+1} \right)^2 = 0 \ , \end{equation}

\noindent where $d = 15$ in this case. These values are quickly found by numerically 
minimizing the function on 
the left-hand side. In practice, the number of terms can be taken to be one 
more than the number of parameters to be determined, and the full SIC condition 
can be checked afterwards. When the precision is high enough the exact phase factors 
$e^{i\sigma}$ can be determined using Mathematica's ``RootApproximant'' command. In 
this case we find that there are three choices of the so far undetermined 
phase factor that give a SIC, namely 

\begin{equation} e^{i\sigma } = 
\omega_3^nP_5^{\frac{1}{3}} \ , \hspace{7mm} P_5 = - \frac{4}{5} - \frac{3i}{5} 
\ , \hspace{5mm} n \in \{ 0,1,2\} \ . \end{equation}

\noindent Indeed, adding the phase factor $P_5^{\frac{1}{3}}$ to the generators of the 
number field holding the SIC 5a gives the number field holding the SIC 15d \cite{SiSi}. 

Note that the symmetry of the 15 dimensional SIC that we constructed is of order 6. 
Its symmetry group is generated by the unitaries corresponding to the symplectic 
matrices

\begin{equation} \mathcal{Z}_{15} = \left( \begin{array}{rr} 1 & 0 \\ 1 & 1 \end{array}\right)_3\times 
\left( \begin{array}{rr} 0 & 2 \\ 2 & -1 \end{array} \right)_5 \sim 
\left( \begin{array}{rr} 10 & 9 \\ 1 & 4 \end{array} \right)_{15} \label{zaunerglobal} 
\end{equation}

\begin{equation} \mathcal{S}_{15} = \left( \begin{array}{rr} -1 & 0 \\ 0 & -1 \end{array} 
\right)_3\times \left( \begin{array}{rr} 1 & 0 \\ 0 & 1 \end{array} \right)_5 \sim 
\left( \begin{array}{rr} 11 & 0 \\ 0 & 11 \end{array} \right)_{15} \ . 
\label{scottglobal} \end{equation}

\noindent In fact the symmetry groups of the aligned SICs grow with a factor of 
two for each rung of the ladder \cite{aligned, AD}. That SICs with these symmetries 
appear in dimensions of the form $d(d-2)$ was first noticed by Scott and Grassl 
\cite{Scott, Andrew}. For later convenience 
we also gave the symplectic matrices obtained from the Chinese remainder theorem when we 
express the Hilbert space globally as ${\bf C}^{15}$ and use the standard 
representation of the Weyl--Heisenberg group there (see the Appendix). 

It is interesting to count the number of aligned SICs obtained. In dimension five the 
symplectic group contains ten distinct Zauner subgroups of order 3, and each of them 
have two eigenspaces containing four SIC vectors each, so in total there are 80 distinct 
but equivalent SICs in the Clifford orbit labelled 5a. In ${\bf C}^3$ there are four 
distinct Zauner subgroups, which (given that each one has two generators) means that 
there are $10\cdot 4 \cdot 2 = 80$ different Zauner subgroups in dimension 15. They 
are generated by matrices of the form $\mathcal{Z}_3\times 
\mathcal{Z}_5$. Having fixed the Zauner matrix $\mathcal{Z}_5$ we have $4\cdot 2$ Zauner 
matrices $\mathcal{Z}_3$ to choose from, and eight SIC fiducials to start from. For each of the 
latter we have checked that the proto-SIC vector family contains a SIC vector in 
${\bf C}^{15}$ for exactly four choices of $\mathcal{Z}_3$. If a given $\mathcal{Z}_3$ works 
its square does not. Thus there are four `empty branches' consisting of families of 
proto-SICs with the expected symmetry but without any SIC vectors in them. Keeping in 
mind that a proto-SIC family contains either three or no SICs we end up with 
$10\cdot 8 \cdot 4 \cdot 3 = 960$ aligned SICs in dimension 15, which is precisely 
the number of distinct but equivalent SICs in the Clifford orbit 15d \cite{Marcus}. 

It is also interesting to consider the triplets of aligned SIC fiducial vectors somewhat 
further. The family of proto-SIC vectors that we defined in equation (\ref{proto15}) defines 
a closed curve in the set of maximally entangled states in ${\bf C}^2\otimes {\bf C}^2 
\subset {\bf C}^{15}$. Using the Fubini--Study metric to define the geometry we find that 
it is a circle of maximal length, and hence it forms the equator of an embedded Bloch 
sphere (or complex projective line, for readers who prefer this language). On this 
equator there are three equidistant points that represent SIC fiducial 
states. In fact they form an orbit under $U_{\mathcal{Z}^{\prime \prime}}^{(3)}
\otimes {\bf 1}_5$. 

\section{Two more examples and a parameter count}\label{sec:moreexamples}

\

\noindent We move on to $d_0 = 7$. There are two inequivalent SICs in this dimension, and 
we choose to work with the one that has aquired the label 7b \cite{Scott}. This time it 
will pay to spend some thought on Step {\bf 1} of the recipe (in Section \ref{sec:recipe}). Since the dimension 
equals 1 modulo 3 there exists a diagonal Zauner matrix, namely 

\begin{equation} \mathcal{Z} = \left( \begin{array}{cc} 2 & 0 \\ 0 & 4 \end{array} \right) 
\ . \end{equation}

\noindent Then $\mathcal{Z}^\prime = \mathcal{Z}$. More importantly the unitary 
operator $U_{\mathcal{Z}}$ is a permutation matrix \cite{Marcus}. The large eigenspace 
(of dimension three) contains four distinct SIC fiducial vectors, two of which are 
real due to an extra anti-unitary symmetry that is present in this case. We choose 
to work with one of these, and calculate the resulting $P_\theta$. We choose 
$\mathcal{Z}^{\prime \prime}$ to be the standard Zauner matrix, go through 
the first three steps of the recipe, and obtain the eigenbases

\begin{equation} \begin{array}{l} |f_0\rangle = |1\rangle_5 \\ |f_1\rangle = 
|\omega_3\rangle_5 \\ |f_2\rangle = |\omega^2_3 \rangle_5 \end{array} \hspace{8mm} 
\begin{array}{l} |e_0\rangle = |1\rangle_7 \\ |e_1\rangle = 
|\omega_3\rangle_7 \\ |e_2\rangle = |\omega^2_3 \rangle_7 \ . \end{array}\end{equation}

\noindent From these basis vectors we want to construct maximally entangled states 
that are eigenvectors of $U_{\mathcal{ Z}^{\prime \prime}}^{(5)}\otimes U_{\mathcal{Z}}^{(7)}$. 
This can be done in more than one way, but we expect (in fact we know \cite{aligned}) 
that there is an aligned SIC in the smallest of the three eigenspaces of this operator. 
We have chosen the phase factors of the Zauner unitaries in dimensions 
$d$ and $d-2$ in conformity with the conventions of Table \ref{tab:spectra}, and 
are thus led to propose the proto-SIC family 

\begin{equation} |\Psi_{\bf 0} (\sigma_1,\sigma_2)\rangle = \frac{1}{\sqrt{3}} 
\left( |f_0\rangle |e_0\rangle + e^{i\sigma }|f_1\rangle |e_2\rangle 
+ e^{-i\sigma}|f_2\rangle |e_1\rangle \right) \ . \label{proto35} \end{equation}

\noindent We use only one free phase factor for the proto-SIC because the anti-unitary 
symmetry of the SIC 7b is inherited by the aligned SIC. A numerical search reveals 
that there are three mutually orthogonal SIC fiducial vectors hidden here. They are 
obtained by letting the sixth power $(e^{i\sigma})^6$ of the phase factor be a root 
of the minimal polynomial 

\begin{eqnarray} p(t) = 4375t^8 - 35000t^7 + 19222300t^6 + 70190980t^5 + 102366979t^4 
\nonumber \\ \\ 
+ 70190980t^3 + 19222300t^2 - 35000 + 4375 \ . \nonumber \end{eqnarray}

\noindent This is no more complicated than it has to be and can be dealt with if 
one wants an exact solution, in this case for the SIC that Scott and Grassl labelled 
as 35j \cite{Scott}. 

So far we did not encounter any degenerate eigenspaces when we defined the bases in 
the factor Hilbert spaces. When $d_0 = 9$ we do. We choose the standard Zauner matrix 
for the dimension 9 factor and a diagonal Zauner matrix for the other factor. At the 
end of Step {\bf 3} we arrive at the bases 

\begin{equation} \begin{array}{l} |f_0\rangle = |1a\rangle_7 \\ |f_1\rangle = 
|1b\rangle_7 \\ |f_2\rangle = |\omega_3^2\rangle_7 \\ |f_3\rangle = 
|\omega_3\rangle_7 \end{array} \hspace{8mm} 
\begin{array}{l} |e_0\rangle = |\omega_3a\rangle_9 \\ |e_1\rangle = 
|\omega_3b\rangle_9 \\ |e_2\rangle = |\omega_3^2\rangle_9 \\ |e_3\rangle = 
|1\rangle_9 \ . \end{array}\end{equation}

\noindent We must now form a maximally entangled state that is an eigenvector of 
the Zauner unitary $U^{(7)}_{\mathcal{Z}^{\prime \prime}}\otimes 
U^{(9)}_{\mathcal{Z}^\prime}$. Moreover we expect the aligned SIC to lie in the 
largest of the three eigenspaces, which in this case (in disagreement with the conventions 
adopted in Table \ref{tab:spectra}) means that the eigenvalue should be $\omega_3$. 
To deal with the degenerate 
eigenspace we recall equation (\ref{snarjelse}), and introduce an $SU(2)$ matrix 

\begin{equation} U = \left( \begin{array}{cc} \cos{\frac{\theta}{2}}e^{i\sigma_0} 
& \sin{\frac{\theta}{2}}e^{i\sigma_1} \\ -\sin{\frac{\theta}{2}}e^{-i\sigma_1} 
& \cos{\frac{\theta}{2}}e^{-i\sigma_0} \end{array} \right) \ . \end{equation}

\noindent A family of proto-SIC vectors invariant under the Zauner unitary 
$U_{\mathcal{Z}^{\prime \prime}}^{(7)}\otimes U_{\mathcal{Z}^\prime}^{(9)}$ is therefore 
given by 

\begin{eqnarray} |\Psi_{\bf 0} \rangle = \frac{1}{\sqrt{4}} 
\Big( \cos{{\tiny \frac{\theta}{2}}}e^{i\sigma_0}|f_0\rangle |e_0\rangle + 
\sin{\frac{\theta}{2}}e^{-i\sigma_1}|f_0\rangle |e_1\rangle - \hspace{12mm} \nonumber \\ \\ 
-\sin{\frac{\theta}{2}}e^{-i\sigma_1}|f_1\rangle |e_0\rangle + \cos{\frac{\theta}{2}}e^{-i\sigma_0}|f_1\rangle |e_1\rangle 
+ e^{i\sigma_2}|f_2\rangle |e_2\rangle + e^{i\sigma_3}|f_3\rangle |e_3\rangle \Big) 
\ . \nonumber \end{eqnarray}

\noindent There are five real parameters in this expression. We thus have a function 
of five variables to minimize, but Mathematica's ``NMinimize command'' handles this 
easily. In this way we identify an aligned SIC. Starting from the SIC labelled 9a by 
Scott and Grassl \cite{Scott} we end up with the SIC that they label 63b. It was originally 
found by a numerical search through the 22 complex dimensional Zauner subspace in 
dimension 63. 

It is clear that numerical searches for aligned SICs can be greatly facilitated 
by our recipe. Let us count the number of free parameters that need to be fixed, assuming 
that the SIC from which we start in dimension $d_0$ has only the Zauner symmetry. The 
details depend a little on the value of $d_0$ modulo 3, and for definiteness we assume 
that $d_0 = 6k+3$. The dimension of the Zauner subspaces will be $(2k+2,2k+1,2k)$ in 
${\bf C}^{d_0}$ and $(2k+1,2k,2k)$ in ${\bf C}^{d_0-2}$. When we restrict ourselves 
to the parity eigenspaces the dimensions drop to $(k+1,k,k)$ in both cases. 
A maximally entangled proto-SIC sitting in the largest Zauner subspace of 
${\bf C}^{d_0-2}\otimes {\bf C}^{d_0}$ will then be given by a block diagonal 
unitary matrix with blocks of size $k+1$, $k$, and $k$. Subtracting an overall 
phase we find that the number of real parameters that must be fixed in order 
to obtain a SIC is 

\begin{equation} (k+1)^2 + k^2 + k^2 -1 = 3k^2 + 2k \ . \end{equation}

\noindent The complex dimension of the relevant Zauner subspace in 
${\bf C}^{d_0(d_0-2)}$ is $12k^2+8k +2$, so this is an improvement 
even though the quadratic dependence on $k$ is still there. For other ways 
to reduce the dimension of the search space see ref. \cite{Dardosjalv}. 

Going beyond this 
reduction in the number of parameters presumably needs ideas from number theory. 

\section{Higher rungs of the ladder}\label{sec:rungs}

\

\noindent The complexity of the calculations, and the size of the proto-SIC family 
that we will have to search through in step {\bf 5} of the recipe (in Section \ref{sec:recipe}), evidently grows 
as we try to reach beyond the first rung of the laddes. Fortunately this problem 
is somewhat mitigated by the extra symmetry that we pick up each time we reach 
a new rung on a ladder. As our example we choose the ladder $5 \rightarrow 15 
\rightarrow 195 \rightarrow 37635 \rightarrow \dots $. The first three entries are 
known in exact form already \cite{SiSi}. Let us see how we can obtain a SIC 
in dimension 195, given the $d_1 = 15$ SIC that we constructed in Section \ref{sec:example}. 
To avoid modifications of the representation given in Section 4 we express the 
latter in `global' form. 

The SIC we are about to construct will have a symmetry group of order 12, generated by 

\begin{equation} \mathcal{Z}_{195} = \left( \begin{array}{rr} 3 & 0 \\ 0 & 9 \end{array} 
\right)_{13}\times \left( \begin{array}{rr} 10 & 9 \\ 1 & 4 \end{array} \right)_{15} 
\end{equation}

\begin{equation} \mathcal{S}_{195} = \left( \begin{array}{rr} 5 & 0 \\ 0 & -5 \end{array} 
\right)_{13} \times \left( \begin{array}{rr} 11 & 0 \\ 0 & 11 \end{array} \right)_{15} \ . 
\end{equation}

\noindent The right hand factors are as given in equations (\ref{zaunerglobal}) and (\ref{scottglobal}). The 
Zauner matrix in the ${\bf C}^{13}$ factor was chosen so that 
its unitary representative is a permutation matrix. But note that 

\begin{equation} \mathcal{S}^2_{195} = \left( \begin{array}{rr} -1 & 0 \\ 0 & -1 \end{array} 
\right)_{13} \times \left( \begin{array}{rr} 1 & 0 \\ 0 & 1 \end{array} \right)_{15} \ . 
\end{equation}

\noindent The parity matrix appears in the left hand factor, and gives rise to the extra 
symmetry of the aligned SIC. This works only because $-1$ is a quadratic residue 
modulo 13. At the next rung of the ladder $-1$ has to be a quartic residue modulo 193, 
and so on. This part of the construction is actually known to work all the way up the 
ladder \cite{IBGM}, but we leave this aside here. 

We define the two $(15-1)/2$-dimensional subspaces as the positive parity eigenspace 
in ${\bf C}^{13}$, and the negative $\theta$-parity eigenspace in ${\bf C}^{15}$. We 
span these eigenspaces with bases consisting of eigenvectors of the relevant factors 
of the unitaries that generate the symmetry group. We end up with the preferred basis

\begin{equation} \begin{array}{l} |f_0\rangle = |1a\rangle_{13} \\ |f_1\rangle = 
|1b\rangle_{13} \\ |f_2\rangle = |\omega_6 \rangle_{13} 
\\ |f_3\rangle = |\omega_6^2 \rangle_{13} \\ |f_4\rangle = |\omega_6^3 \rangle_{13} 
\\ |f_5\rangle = |\omega_6^4 \rangle_{13} \\ |f_6\rangle = |\omega_6^5 \rangle_{13} 
\end{array} \hspace{12mm} 
\begin{array}{l} |e_0\rangle = |1a\rangle_{15} \\ |e_1\rangle = 
|1b\rangle_{15} \\ |e_2\rangle = |\omega_6 \rangle_{15} \\ |e_3\rangle = 
|\omega_6^2 \rangle_{15} \\ |e_4\rangle = |\omega_6^3 \rangle_{15} \\ 
|e_5\rangle = |\omega_6^4 \rangle_{15} 
\\ |e_6\rangle = |\omega_6^5 \rangle_{15} \ . \end{array}\end{equation}

\noindent A family of proto-SICs is now formed as maximally entangled states having 
the appropriate symmetry (eigenvalue 1, in this case). It is given by 

\begin{eqnarray} |\Psi_{\bf 0} \rangle = \frac{1}{\sqrt{7}} 
\Big( \cos{{\tiny \frac{\theta}{2}}}e^{i\sigma_0}|f_0\rangle |e_0\rangle + 
\sin{\frac{\theta}{2}}e^{-i\sigma_1}|f_0\rangle |e_1\rangle - \hspace{12mm} \nonumber \\ 
\nonumber \\ 
-\sin{\frac{\theta}{2}}e^{-i\sigma_1}|f_1\rangle |e_0\rangle + \cos{\frac{\theta}{2}}e^{-i\sigma_0}|f_1\rangle |e_1\rangle + 
\hspace{12mm} \\ \nonumber \\ 
+ e^{i\sigma_2}|f_2\rangle |e_6\rangle + e^{i\sigma_3}|f_3\rangle |e_5\rangle 
+ e^{i\sigma_3}|f_4\rangle |e_4\rangle + e^{i\sigma_5}|f_5\rangle |e_3\rangle 
+ e^{i\sigma_6}|f_6\rangle |e_2\rangle \Big) 
\ . \nonumber \end{eqnarray}

\noindent It is an eight real dimensional 
family, small enough so that we can use standard Mathematica routines for the 
numerical search that determines the 195 dimensional SIC. By construction it has a 
symmetry of order 12, and is equivalent to the SIC labelled 195d \cite{aligned}. 
We again used standard Mathematica routines to determine that $\cos{\theta/2} = 
3/\sqrt{13}$ and that the sixth powers of the phase 
factors can be obtained by finding the roots of seven polynomials of degree 32 and 
large coefficients.\footnote{To reach this conclusion we increased the precision of the 
numerical vector from 200 to 3400 digits using a Mathematica file kindly provided by Marcus 
Appleby. This calculation took 773 seconds.} This is encouraging given that the ray 
class field that houses the SIC has degree $2^8\cdot 3^3$. The solution is easily brought 
to the more elegant form given in ref. \cite{SiSi}. 

The logic remains the same as we continue up the ladder. To reach the next rung 
we consider ${\bf C}^{193}\otimes {\bf C}^{195}$. The symmetry operators 
in the ${\bf C}^{193}$ factor can again be chosen as permutation matrices, and the 
family of proto-SICs with the expected symmetry has 98 real parameters. However, 
here we encounter the problem that the present authors have no expertise 
in numerical optimization. We have not solved it, and hence we have not 
found the $d = 37635$ SIC. This is a pity, because there are some reasons to believe 
that all the SICs on this ladder can be written in an especially appealing exact form 
\cite{IBGM, SiSi}. If this is true we would need only modest numerical precision in order 
to obtain the exact solution through an integer relation algorithm, and the result 
might suggest how to go even higher. 

\section{Summary}\label{sec:summary}

\

\noindent We believe that we have clarified how aligned SICs, as defined in ref. 
\cite{aligned}, arise. By a suitable arrangement of known results we have shown that 
the set of Naimark complements of the equiangular tight frame appearing in Theorem 
\ref{Theorem1}, which is parametrized by the group manifold of $SU((d-1)/2)$, can 
be lifted to ${\bf C}^{d(d-2)}$ where they form maximally entangled states in a subspace 
that is defined by numbers entering the SIC in ${\bf C}^d$. This leads to a straightforward 
calculational procedure for producing a family of 
proto-SIC vectors in dimension $d(d-2)$, given a SIC in dimension $d$ 
to start with. By definition, a proto-SIC vector obeys the alignment conditions 
(\ref{aligned1}) and (\ref{aligned2}), but it is not necessarily a fiducial vector 
for a SIC. We have also explained how the ambiguities that 
we have encountered can be partly removed by imposing symmetry conditions. Still 
the full SIC condition eludes us, so we have not proved that aligned SICs must exist. 
What we do have is a remarkably convenient way to conduct numerical searches for 
aligned SICs. Even so we have not calculated any new SICs, because the easy targets 
have already been calculated with other methods. For $d = 195\cdot 193 = 37635$ we 
would have to optimize a function of 98 real variables. While this is not an 
insuperable task, it is insuperable for the present authors. 

We end by echoing a remark of Appleby's \cite{Marcus}: The crucial discovery may lie 
just round the corner. 

\vspace{1cm}

\noindent \underline{Acknowledgements}: We thank Irina Dumitru for discussions in the 
course of this work. We also thank Markus Grassl, Danylo Yakymenko, and Ole Sönnerborn 
(one of the authors of ref. \cite{AD}) for very useful comments 
on a draft. BS acknowledges the financial support from the Knut and Alice Wallenberg 
Foundation through the Wallenberg Centre for Quantum Technology. 

\

\appendix

\section{Useful formulas}\label{sec:App}

\

\noindent For the convenience of the reader we collect some useful formulas here. 
For the displacement operators we use the representation

\begin{equation} (D_{i,j})_{r,s} = \tau^{ij + 2js}\delta_{r,s+i} \ . \end{equation} 

\noindent For the symplectic unitaries the requirement that $U_FD_{\bf p}U_F^{-1} = 
D_{F{\bf p}}$ implies 

\begin{equation} F = \left( \begin{array}{cc} \alpha & \beta \\ \gamma & \delta 
\end{array} \right)_d \hspace{5mm} \Rightarrow \hspace{5mm} (U_F)_{r,s} = 
\frac{e^{i\varphi}}{\sqrt{d}} \tau^{\beta^{-1}(\delta r^2 - 2rs + \alpha s^2)} 
\end{equation}

\noindent whenever the integer $\beta$ is invertible modulo $d$. If not, the symplectic 
matrix can be written as a product of two matrices with an invertible integer in 
the upper right hand corner, and the symplectic unitary as a product of the two 
corresponding symplectic unitaries. The overall phase factor is chosen so that Table \ref{tab:spectra} applies. For the Chinese remaindering we use 

\begin{equation} \left( \begin{array}{cc} \alpha & \beta \\ \gamma & \delta 
\end{array} \right)_{d(d-2)} \sim \left( \begin{array}{cc} \alpha & \kappa^{-1}\beta \\ 
\kappa \gamma & \delta \end{array} \right)_{d-2}\times \left( \begin{array}{cc} 
\alpha & \kappa^{-1} \beta \\ \kappa \gamma & \delta \end{array} \right)_d
\end{equation}

\noindent where $\kappa = (d-1)/2$. 

\

\end{document}